# Research Integrity and Academic Authority in the Age of Artificial Intelligence: From Discovery to Curation?

Simon Chesterman* and Loy Hui Chieh[†]

*Artificial intelligence is reshaping the organisation and practice of research in ways that extend far beyond gains in productivity. AI systems now accelerate discovery, reorganize scholarly labour, and mediate access to expanding scientific literatures. At the same time, generative models capable of producing text, images, and data at scale introduce new epistemic and institutional vulnerabilities. They exacerbate challenges of reproducibility, blur lines of authorship and accountability, and place unprecedented pressure on peer review and editorial systems. These risks coincide with a deeper political–economic shift: the centre of gravity in AI research has moved decisively from universities to private laboratories with privileged access to data, compute, and engineering talent. As frontier models become increasingly proprietary and opaque, universities face growing difficulty interrogating, reproducing, or contesting the systems on which scientific inquiry increasingly depends.*

---

* Simon Chesterman is David Marshall Professor and Vice Provost (Educational Innovation) at the National University of Singapore, where he is also the founding Dean of NUS College. He serves as Senior Director of AI Governance at AI Singapore and Editor of the *Asian Journal of International Law*.

[†] Loy Hui Chieh is an Associate Professor of Philosophy at the National University of Singapore, concurrently also Vice Dean (Academic Affairs) and Master of NUS College. He also chairs the NUS University Policy Workgroup for AI in Teaching and Learning.

Many thanks to Fakhar Abbas, Kan Min-Yen, Ben Leong, Hakim Norhashim, Eka Nugraha Putra, Araz Taeihagh, Chen Tsuhan, Melvin Yap, Audrey Yue, and many others for rich discussions on the material presented here. Earlier iterations of this work have benefited from discussions at Lingnan University, Nanyang Technological University, the National University of Singapore, Peking University, and Shanghai Jiao Tong University. Invaluable research assistance was provided by He Yiyang and Shambhavi Mehra. Errors, omissions, and hallucinations are attributable to the authors alone.

*This article argues that these developments challenge research integrity and erode traditional bases of academic authority, understood as the institutional capacity to render knowledge credible, contestable, and independent of concentrated power. Rather than competing with corporate laboratories at the technological frontier, universities can sustain their legitimacy by strengthening roles that cannot be readily automated or commercialized: exercising judgement over research quality in an environment saturated with synthetic outputs; curating the provenance, transparency, and reproducibility of knowledge; and acting as ethical and epistemic counterweights to private interests. In an era of informational abundance, the future authority of universities lies less in maximizing discovery alone than in sustaining the institutional conditions under which knowledge can be trusted and publicly valued.*

# Introduction

When the Nobel Prize in Chemistry for 2024 was presented by King Carl XVI Gustaf to Demis Hassabis and John Jumper of DeepMind, together with David Baker of the University of Washington, the applause was warm and sustained. Their work on computational protein design and protein structure prediction solved a problem that had frustrated biochemists for half a century, opening up fields of research and the possibility of designing entirely new proteins (Åqvist, 2024; Senior, et al., 2020).

For universities, the moment was quietly disorienting. The world's highest scientific honour — long the emblem of academic distinction — had been awarded for research conceived, executed, and funded inside a corporate lab. Watching from faculty offices and seminar rooms, some might have experienced a tremor beneath the ceremony: a recognition that the frontier of discovery was shifting, perhaps permanently, away from academe.

Universities have long asserted their value and legitimacy as institutions that produce research as well as education. Since the emergence of the research university in the nineteenth century (Levine, 2021), this dual role has been central to their place in modern society. In terms of value, research can complement education: it refreshes what is taught, sustains the authority of the professoriate, and underpins the claim that universities are not simply transmitters of wisdom but creators of new ideas and capacities. In terms of legitimacy, the knowledge and expertise of academics long granted them a privileged place in society: their opinions given varying degrees of deference; their independence protected by the conventions of tenure and academic freedom (Altbach, 2011).

Artificial intelligence has the potential to accelerate the research mission of universities in profound ways.[1] By processing vast amounts of data, AI systems are enabling discoveries once thought impractical or impossible — from protein folding to analysing the history and authorship of the Dead Sea Scrolls (Popović, et al., 2025). Such advances demonstrate the extraordinary potential of AI as a research tool (Renkema and Tursunbayeva, 2024). Even as the frontier of discovery is widening, however, the locus of AI research itself is shifting decisively from universities to industry. The large language models at the heart of generative AI may have been conceptualized in academic laboratories, but the data, compute, and talent now required place the cutting edge firmly in the hands of private firms. To the extent that future innovation will rely on such technologies, it raises a deeper challenge: if universities are no longer central to knowledge production, on what basis can they continue to claim legitimacy as independent centres of learning, rather than merely service providers in a commercial research ecosystem?

The field is fast-moving, and there is evidence of academic researchers unfamiliar with the technology offering teaching moments for a wider audience. After provocateurs tried to list 'ChatGPT' as a co-author of several articles in academic journals in 2022 and 2023 (Stokel-Walker, 2023), most of these were withdrawn and new rules were adopted to prohibit (for the time being) AI from being recognized as an author.[2] That did not stop others from ploughing ahead, with a series of embarrassing publications in 2024 that included text clearly stating that it had been produced by generative AI. A noteworthy example was the very first sentence of an article in a peer-reviewed materials science journal, which began: 'Certainly, here is a possible introduction for your topic' (Zhang, et al., 2024).

There is a rich literature on the nature and function of universities, much of it illuminating, some of it self-indulgent. For our purposes, we connect that literature to wider debates about the political economy of higher education (Münch, 2014; Slaughter and Rhoades, 2004), the sociology of professions (Abbott, 1988; Larson, 1977), and the shifting legitimacy of knowledge institutions in an era of technological disruption and populist scepticism (Collins and Evans, 2007; Oreskes, 2019). In doing so, we draw on analyses that treat universities not only as places of learning, but also as labour-market actors, cultural intermediaries, and contested sites of authority.

---

[1] For the purposes of this article, 'artificial intelligence' (AI) refers broadly to computational techniques — particularly machine learning methods, including deep learning and large language models — that enable systems to perform tasks such as pattern recognition, prediction, generation, and optimization across research workflows, rather than to any claim about general or human-like intelligence.

[2] See, eg, Nature, Science, Elsevier, among others.

'Academic authority', for present purposes, refers to the institutionalized form of epistemic legitimacy through which universities and scholarly communities claim the capacity to produce, evaluate, and validate knowledge independently of political or commercial power. That authority rests not simply on individual expertise or credentials, but on collectively sustained practices — including peer review, methodological transparency, organized scepticism, and professional autonomy — that render knowledge claims trustworthy beyond the individuals who advance them (Merton, 1973; Weber, 1978). Long understood as a defining feature of the modern research university, academic authority is therefore both relational and fragile: it depends on social trust in epistemic practices and on institutional conditions that allow those practices to operate with a degree of independence from external interests (Jasanoff, 2004; Shapin, 1994).

The article first examines how AI is reshaping research practice by functioning as a form of research infrastructure, accelerating discovery and lowering barriers to participation across a range of disciplines. Section two then turns to the epistemic and integrity risks that accompany these developments, including challenges to reproducibility, authorship, peer review, and the reliability of knowledge. These risks are not merely technical or individual in nature, but are embedded in broader institutional incentives and professional norms. Section three situates these dynamics within the political economy of AI research, analysing the growing concentration of data, compute, and talent in private laboratories and elite institutions, and the implications of that concentration for openness, accountability, and the distribution of research capacity. The final section considers how universities might respond to these shifts, arguing that they should not — indeed cannot — compete at the technological frontier, but rather should reposition themselves as evaluators, curators, and ethical stewards of knowledge in an environment that will be increasingly saturated with synthetic information of uncertain quality.

# 1 AI as Research Infrastructure

Universities have long played a key role in the scientific and technological breakthroughs that transformed modern life. Academic laboratories developed the insulin therapy that turned type 1 diabetes from a death sentence into a manageable condition; enabled the discovery and mass production of penicillin, ushering in the antibiotic era; and laid the foundations of the global internet through early ARPANET work at UCLA, UC Santa Barbara, Utah, and the Stanford Research Institute (Lukasik, 2011). University-based researchers created the first viable lithium-ion battery cathode, powering today's portable electronics and electric vehicles, and pioneered the mRNA technologies later deployed at scale during the COVID-19 pandemic. Even the PageRank algorithm that drove Google's rise emerged from graduate work at Stanford, which went on to patent the design (Page, et al., 1999).

These achievements exemplify the central role that universities played as engines of curiosity-driven inquiry — a role now being reshaped by the shifting locus of innovation and the growing dominance of private-sector AI research. Universities are not immune to the pressures to commercialize, but in the past tended to regard this as a linear model in which a scientist made a discovery, a technology transfer office (or equivalent) determined whether to seek a patent, which might then be licensed to a company or spun off in the form of a startup (Bradley, et al., 2013; Carlsson and Fridh, 2002). Around the world, there is now greater pressure to add innovation to education and research as part of the core mission of universities (Menter, 2024).

The suite of technologies embraced by the term AI offers the most significant expansion of the research toolkit since the advent of modern computation, with the potential to accelerate discovery, lower barriers to participation, and reshape the everyday practices of academic work. Much as earlier generations of researchers embraced statistics or simulation modelling as methodological revolutions, today's researchers are increasingly turning to machine learning systems not merely as labour-saving devices but as collaborators of a sort: tools capable of parsing vast datasets, generating hypotheses, and exploring problem-spaces that were previously inaccessible. At its most optimistic, AI promises something close to an 'invention of a method of invention': a set of techniques that can spur innovation across multiple fields, with an impact potentially exceeding that of any single breakthrough (Cockburn, et al., 2019: p. 116).

## 1.1 Accelerating Discovery

The most visible example of accelerated discovery is the dramatic progress in protein structure prediction, culminating in the 2024 Nobel Prize. Problems that might once have consumed decades of experimental work can now be approached computationally, guiding laboratory effort and reducing the cost of exploration. Similar advances are emerging in drug discovery more broadly. The field of antibiotics, long mired in stagnation — twenty new classes of antibiotics were discovered by 1962 but only a handful since then, the last arguably in 1987 (Mohr, 2016; Walsh, 2003) — has been revitalized by deep-learning models able to screen millions of candidate molecules. Systems using such techniques have already identified compounds with entirely novel mechanisms of action, raising the real possibility of rebooting an area of research that had been considered commercially unviable and methodologically exhausted (Stokes, et al., 2020).

Comparable transformations are underway in materials science, where machine-learning models now sift through immense combinatorial spaces of potential compounds to identify promising candidates for batteries, catalysts, or photovoltaics (Morgan and Jacobs, 2020). Experimentalists benefit from the ability to prioritize only the most viable options, reducing the burden of trial-and-error approaches. Even within more theoretical domains, AI is beginning to show signs of contributing to problem-solving, such as assisting in the construction of formal proofs or generating conjectures that researchers can later refine (Beel, et al., 2025; Polu and Sutskever, 2020). These developments do not yet replace the need for human insight, but they do expand the frontier of problems that can realistically be tackled.

A more mundane but no less important aspect of discovery lies in the ability of AI to process the scientific literature itself. Researchers across all fields struggle to keep pace with the volume of published work; even narrow sub-disciplines now generate thousands of articles annually. Large language models can scan, classify, and summarize this material at a scale no human could match, pointing to overlooked connections or emerging trends. In interdisciplinary areas where researchers must integrate insights from fields in which they were not trained — computational biology or environmental data science, for example — such tools can act as translators or guides, flattening the learning curves that once hindered collaboration (Mammides and Papadopoulos, 2024). Used well, this may lead not to homogenization but to deeper cross-fertilization among disciplines.

## 1.2  Lowering the Barriers to Entry

These capabilities also raise the possibility of democratization of research. Free or low-cost AI systems increasingly offer sophisticated analytical support for tasks such as coding, statistical modelling, and even the drafting of preliminary experimental designs. For researchers in less-resourced institutions or in countries with limited access to advanced infrastructure, such tools may offer a level of capability previously restricted to elite laboratories. In theory, this could broaden participation in global research networks and enable new voices to contribute to cutting-edge problems (Nji, et al., 2025). In practice, that promise is tempered by real constraints: access to reliable internet and electricity cannot be assumed; linguistic disparities persist, with most systems performing best in English; and many AI tools still rely on proprietary models hosted by commercial providers. Nevertheless, the tools have genuine democratizing potential, even if current access is uneven. This paradox — lowering of some barriers even as proprietary models raise new ones — will be considered further in section three.

Laboratory automation offers another avenue for expanding research capacity and widening access. Advances in robotics, coupled with AI systems capable of designing and interpreting experiments, have led to the emergence of 'self-driving laboratories', particularly in chemistry and materials science (Tom, et al., 2024). Such systems allow experiments to be planned, executed, analysed, and iteratively refined with limited human intervention. This potentially offers small laboratories the ability to undertake high-throughput experimentation of a kind once reserved for major research centres. Although the technology remains in its early stages, and many disciplines still require substantial hands-on training and tacit knowledge that cannot be automated (Collins, 2010), these developments hint at a future in which physical experimentation can be scaled through computation rather than lab space. For universities, this may eventually alter the economics of laboratory research, reducing the role of brute-force capability and placing greater value on creativity and cross-disciplinary problem framing, and potentially affecting hiring decisions and even the design and construction of laboratories themselves.

## 1.3  The Practice of Academic Life

The opportunities are not confined to the laboratory bench. AI is also beginning to transform the everyday administration associated with academic research. Preparing grant applications, drafting ethics proposals, and compiling activity reports are among the most time-consuming — and least rewarding — duties of many researchers. Generative AI can streamline such tasks, allowing academics to focus more of their energies on conceptual and experimental work. Even in writing up research findings, AI can help organize material, improve clarity, and generate alternative formulations for complex arguments. Recent studies suggest that such tools can enhance writing quality and reduce the cognitive load associated with drafting (Haber, et al., 2025; Silva, et al., 2025). This is not without controversy: in the humanities and interpretive social sciences, the act of writing is often inseparable from the intellectual work itself, and there are legitimate concerns about eroding the craft of scholarly expression (Jeon, et al., 2025). In many STEM fields,[3] by contrast, writing is regarded more as a form of technical communication, and assistance from AI is welcomed as a means of increasing productivity (Chubb, et al., 2022). These disciplinary differences mirror broader tensions about the appropriate use of generative models in creative tasks, but they do not diminish the potential for AI to reduce the burden of academic drudgery.

Taken together, the opportunities presented by AI run deeper than any single breakthrough might suggest. They point to a shift in the research ecosystem as a whole — widening the aperture of what problems can be addressed, who can address them, and how research is conducted day to day. The challenge, as explored in subsequent sections, is how to seize these opportunities while navigating the new ethical and practical risks that they bring, and at a time when the centre of gravity in AI research itself has moved decisively away from academia and towards the private sector.

## 2  Epistemic and Integrity Risks in AI-Driven Research

The near-term ethical concerns arising from increased reliance on artificial intelligence in research are, by now, familiar. Generative AI in particular is already influencing virtually every phase of the research cycle — shaping hypotheses, designing experiments, analysing data, visualizing and interpreting findings, drafting manuscripts, and even participating, directly or indirectly, in the peer-review process (Vasconcelos and Marušić, 2025). In many respects, this is an extension of longstanding dilemmas about the appropriate use of computational tools in science and scholarship. Yet the speed, scale, and opacity of contemporary AI systems give these issues new weight in the areas of reproducibility, research integrity, and structural risks to the research enterprise as a whole.

---

[3] STEM refers to the disciplines of science, technology, engineering, and mathematics, commonly grouped together to denote fields characterized by formal, technical, and quantitative methods of inquiry.

## 2.1 Reproducibility

The most obvious problems echo earlier debates about bias and unrepresentative data (Oduoye, et al., 2023). AI models trained predominantly on Western, educated, industrialized, rich, and democratic ('WEIRD') populations replicate the same limitations that have long troubled social science datasets based on narrow subject pools (Henrich, et al., 2010). If those earlier concerns were about undue weight placed on WEIRD samples, the contemporary worry is that vast and inscrutable training corpora embed unexamined assumptions at a planetary scale. This creates fertile ground for errors, privacy breaches, and the spread of misinformation. Traditional research ethics and institutional review structures, designed around human subjects and incremental methods, struggle to keep pace with tools capable of generating limitless synthetic data or producing persuasive interpretations that command undue trust (Chen, et al., 2024).

A distinct cluster of risks concerns the reliability and epistemic quality of AI-supported research. Hallucinations — fabricating plausible but false information, confidently expressed — remain an endemic problem, especially when models are pushed beyond, or required to extrapolate from, the distributions represented in their training data. More subtle are issues of data drift or distributional shift, in which models trained on one kind of data degrade when faced with another (Quiñonero-Candela and Neural Information Processing Systems, 2009). Model poisoning is an emerging threat, with adversaries injecting corrupted or misleading samples into training corpora (Yazdinejad, et al., 2024). There is also a growing worry about synthetic feedback loops, in which AI-generated outputs are used to train subsequent models, gradually narrowing the epistemic diversity of the system — an 'AI groupthink' that ossifies prevailing patterns rather than discovering new ones (Shumailov, et al., 2024).

All of this undermines the ability of researchers to reproduce or audit claims for which they remain — in theory at least — responsible.

## 2.2 Research Integrity

These problems, however, gesture towards a deeper threat: the pressure placed by AI on the integrity of the research enterprise itself. Historically, breaches of research integrity have arisen from deliberate human misconduct or a failure to exercise due care — manipulating data, misrepresenting methods, or falsifying evidence. Generative AI transforms these risks both in kind and in scale (Wilson and Burleigh, 2025). Such tools dramatically lower the cost of producing text, images, and synthetic datasets that can obscure poor practice or inflate publication output. In a system already strained by the 'publish or perish' imperative, and in which university rankings amplify incentives to prioritize quantity over quality, AI offers a perfect storm (Watermeyer, et al., 2024). Pressures that previously shaped individual behaviour now operate at the level of institutional strategy, with generative systems enabling a volume of output that outstrips the capacity of peer communities to vet, interpret, or even read it.

These concerns have arisen most clearly in debates about authorship. As noted earlier, mischievous authors briefly tested the limits of editorial policies by listing models such as ChatGPT as co-authors, prompting rushed efforts to craft or clarify editorial policies (Limongi, 2024). Though individual journals differ in emphasis, a consensus has emerged across major publishers that AI systems cannot be so named.

Debates over authorship are merely patches on the larger question of whether any human possesses an intelligible understanding of how a given claim generated through the use of AI was produced, sufficient to lend it credibility and authority based on demonstrated skills and reputation in the field. Many policies therefore go on to stress that the use of such tools must be disclosed transparently, and that human authors remain fully accountable for the accuracy, interpretation, and originality of their work (Blau, et al., 2024; Cheng, et al., 2025). Substantial human contribution remains essential, whether in the conception and design of the work, the analysis and interpretation of data, or (at the very least) the careful crafting, curation, and integration of prompts and outputs (Porsdam Mann, et al., 2024). Parallel concerns arise in peer review: reviewers may not upload confidential manuscripts into external AI systems, and ultimate responsibility for the review rests with humans (Perlis, et al., 2025).

Yet detection is difficult. Some commentators report the use of AI-generated 'white text' embedded invisibly in draft manuscripts to evade scrutiny and trick automated reviewers — posing intriguing ethical questions as such subterfuge by authors would only be effective if reviewers are failing in their own obligations by outsourcing the review process to machines (Gibney, 2025).

At a systemic level, a potentially exponential increase in the number of papers produced with AI assistance risks overwhelming editorial and peer-review capacity, drowning high-quality work in a sea of noise. Some publishers have responded with measures such as limiting the number of daily submissions, mirroring similar steps taken by commercial platforms to curb low-quality, AI-generated content or 'slop'. The danger is not simply that poor-quality research proliferates, but that the signal-to-noise ratio becomes so bad that the collective ability of scholars to discern meaningful contributions is impaired.

Despite these hazards, it is also possible that AI might help bolster or police research integrity. Machine-learning tools are already used to identify duplicated images, statistical anomalies, or text recycled across publications (Bauchner and Rivara, 2024). Pre-publication checks could, in principle, flag ethical concerns, assess data validity, or screen for bias. Such possibilities have prompted researchers to articulate guiding principles for the responsible use of AI in research more generally: transparent disclosure and attribution of AI involvement; verification of AI-generated content; documentation of synthetic or AI-derived data; respect for consent and privacy; fairness and equity in access; and continuous monitoring and oversight (Embracing AI with Integrity, 2025). These principles are not novel — they extend long-standing commitments to openness, integrity, and accountability — but they take on new urgency as AI becomes embedded in routine scholarly practice.

## 2.3  Structural and Systemic Risks

These epistemic challenges are compounded by structural and systemic risks to the academic research enterprise as a whole. Intellectual property questions remain unresolved: many AI systems are trained on copyrighted or pirated material, raising concerns about the legality of inputs and the ownership of outputs (Chesterman, 2025a). The environmental cost of large-scale training runs is increasingly apparent, with significant energy consumption and water usage concentrated in a small number of private-sector facilities (Crawford, 2021). Researchers also warn of deskilling. If routine analytic or conceptual tasks are delegated to models, the expertise that underpins professional judgment may erode — a phenomenon anticipated in other domains and captured in Susskind and Susskind's account of the changing 'grand bargain' of the professions (Susskind and Susskind, 2015: pp. 9–45). At stake is nothing less than the authority of universities as arbiters of knowledge. If institutions accept AI outputs uncritically, outsourcing not only labour but epistemic judgment, then their legitimacy — already under pressure — may be further undermined.

All of this raises acute questions about the future of the research profession. The worry is that if AI automates much of the 'junior' work — literature reviews, data cleaning, basic analysis — future researchers may never acquire the foundational craft knowledge upon which the capability for independent scholarship is built (Hutson and He, 2024). Will academics increasingly adopt the role of principal investigators orchestrating fleets of AI assistants, much as architects today rarely draw by hand and rely on computer-aided design (CAD)? And what becomes of meaning-making in disciplines where interpretation itself is central? Experiments with chatbots mimicking famous philosophers, for example, raise questions about whether AI deepens or trivializes engagement with complex ideas (Kasneci, et al., 2023; Stoljar and Zhang, 2024). These tensions, though still emergent, point to choices about what kind of research culture universities wish to cultivate.

In some disciplines, notably mathematics, the value of solving hard problems lies not primarily in arriving at a final answer, but in the new understanding, methods, and conceptual tools developed in the process. This point has been reiterated recently by Jonathan Gorard in response to speculation about potential AI solutions to the Navier–Stokes equations, but it has a much longer pedigree. The interest of problems such as Fermat's Last Theorem or the Poincaré Conjecture lay precisely in the way they exposed limitations in existing frameworks and, in resolving them, generated new mathematical ideas with wide applicability. The concern is not simply that AI systems might generate solutions more quickly, but that such solutions could bypass the forms of explanation and conceptual restructuring through which mathematical knowledge traditionally advances — a worry long associated with purely mechanical proofs or proof-by-computation, which many mathematicians regard as epistemically unsatisfying even when formally correct (Thurston, 1994; Tymoczko, 1979).

The greatest risk, however, may be irrelevance. If universities neither shape the development of AI nor safeguard the integrity of research conducted with it, they risk ceding authority to actors whose incentives may not align with curiosity-driven inquiry or the public good. This prospect foreshadows the challenges explored in the section four: how universities might remain meaningful contributors to the evolving research landscape rather than spectators to its transformation.

# 3 The Political Economy of AI Research

The siting of significant research outside universities is not new. Prior to the emergence of the modern research university, many of the scientific discoveries that shaped the Scientific Revolution of the sixteenth and seventeenth centuries — and the Enlightenment that followed — took place outside academic institutions. For much of the twentieth century, similarly consequential research occurred in government-run or wholly commercial settings, including AT&T's Bell Labs, the Manhattan Project, and the RAND Corporation. And, as discussed earlier, many foundational discoveries made within universities were subsequently commercialized and scaled by industry.

What makes AI different is its role as a general-purpose technology with broad social, economic, and political consequences. As talent, compute, and data concentrate in a small number of firms, the incentives that shape research are increasingly aligned with profit rather than public interest (Chesterman, 2026). In the short-term, this accelerates deployment but weakens oversight; in the medium-term, it channels effort toward commercially lucrative but socially narrow outcomes; and in the longer-term, it threatens to undermine the capacity of universities to sustain their legitimacy as independent contributors to the knowledge economy.

## 3.1 Investment and Incentives

This is true most obviously of the foundational research on which contemporary AI systems are built. As noted in the introduction, generative AI systems and the intellectual foundations for them first appeared in universities but frontier research in this domain is now dominated by industry. The migration of AI researchers is well-documented (Jurowetzki, et al., 2025). One simple reason is the availability of raw computing power, and on this score, commercial enterprises outmatch even the top research universities. In 2021, the U.S. federal government allocated US$1.5 billion to non-defence academic research into AI; Google spent that much on DeepMind alone. By 2024, 90 percent of notable AI models came from industry rather than academia (AI Index 2025, 2025). In addition to talent and compute, private sector access to data is vastly greater because of a cavalier attitude towards intellectual property that no serious university would allow (Chesterman, 2025a). With projected expansions of data centres and ever more powerful models being deployed, the dominance of the private sector in frontier AI research seems set to continue.

Yet it is also becoming true in other areas of research, which increasingly rely on AI as well as the private sector. The 2024 Nobel Prize for Chemistry is merely the most prominent example. While other companies have played prominent roles in transformative work — Bell Labs, for example, has contributed to at least 9 Nobel Prizes (Georgescu, 2022) — this was almost always in partnership with academic institutions, which by themselves account for something like 70 percent of such awards over the past century (Zhang and Zhang, 2023). Such research also took place in a broader ecosystem of publicly funded basic science, whereas contemporary AI breakthroughs increasingly rely on proprietary data and compute that universities cannot access on comparable terms.

'AI+X' research offers great promise, but there is already evidence of an impact on the focus of investigation in one of the areas adjacent to fundamental AI research itself: safety. Corporate AI research tends to focus on pre-deployment areas, such as model alignment and testing and evaluation, while attention to deployment-stage issues such as model bias is less likely to be supported (Strauss, et al., 2025). More generally, there is evidence of pressure to scale back activities that might slow development and deployment of ever more powerful models, due to concerns about profitability or competition (Chesterman, 2025b). As an expanding set of research fields come to rely on AI, such shifts may become normalized and, eventually, the norm.

The skew reflects the familiar logic of the profit motive and corporate quarterly earning calls. AI systems that optimize advertising revenues, automate customer service, or streamline logistics have clear business cases. Research that addresses unmet social needs — such as rare diseases, long-horizon climate adaptation, or low-resource languages[4] — often does not. The point is not that companies never engage in such work, but that it will always be hard-pressed to compete internally with more obviously monetizable projects. Universities, in principle, should be the institutions that pursue these neglected questions: work that is too speculative, too long-term, or too geographically diffuse to appeal to shareholders. Yet as AI becomes ever more capital-intensive, their ability to do so is constrained by the very resource disparities that prompted the corporate shift in the first place.

## 3.2 The Privatization of the Research Ecosystem

The consequences extend beyond topic selection to the structure of the research ecosystem. Access to frontier models and compute is concentrated in a small number of firms and elite institutions, primarily in the Global North (UN AI Advisory Body, 2024). Researchers in less-resourced universities or in the Global South frequently depend on free tiers of commercial systems, which may be restricted, throttled, or withdrawn at any time. This exacerbates long-standing disparities in who can pose and answer research questions. Efforts such as Masakhane — a grassroots network building natural-language processing tools for African languages through open and collaborative methods — show what community-driven, de-centralized AI research can look like. Such consortia illustrate that collective organization, rather than institutional scale, may be the key determinant of who can still shape research agendas in a corporate-dominated ecosystem. But they remain exceptions, rather than the rule (Agbeyangi and Suleman, 2024).

---

[4] A 'low-resource language' refers to a language for which there is limited digitally available data — such as annotated corpora, lexicons, or training datasets — constraining the development and performance of computational language models.

At the epistemic level, the dominance of a handful of private actors risks fragmenting the commons on which modern science depends. Over the past decade, debates about 'open' and 'closed' AI have often obscured the fact that many ostensibly open models are open in name only. Firms may release code and model weights while withholding training data or evaluation pipelines, offering the appearance of transparency without its substance (Widder, et al., 2024). Such 'open-washing' preserves corporate control but weakens the foundations of scientific scrutiny.

That opacity has predictable consequences. Independent researchers have raised concerns that AI-driven research is contributing to the reproducibility crisis discussed earlier, particularly when the underlying models are proprietary (Gibney, 2022). It is increasingly difficult to verify claims built on closed systems whose training data, architectures, and failure modes remain inaccessible to the wider scientific community (Ball, 2023). In fields where minor variations in data or parameters can radically alter outcomes, such barriers are not merely inconvenient — they undermine the collective capacity of researchers to interrogate and validate knowledge.

A related problem lies in the stratification of research capacity. Though machine-learning scholarship once benefited from the democratizing influence of open-source software and preprint culture, the escalating costs of frontier research have led to a 'compute divide' (Ahmed, et al., 2023). Participation in elite AI conferences now correlates strongly with access to specialized hardware and large-scale compute; institutions without such resources are effectively excluded from the frontier (Lehdonvirta, et al., 2025; AI Index 2025, 2025). This disparity is particularly pronounced among less-resourced universities and in the Global South, where researchers often rely on restricted free-tier or older models. These systemic constraints shape not only the questions researchers can pursue, but also the distribution of who is able to contribute to global scientific debates.

## 3.3 The Erosion of Academic Authority

Taken together, these trends erode the academic authority that universities enjoyed for the past century. Since the rise of the research university model in the nineteenth century, universities have claimed legitimacy not merely as transmitters of knowledge but as institutions capable of interrogating, verifying, and sometimes overturning it. When the tools that increasingly mediate scientific discovery are built and controlled by private firms — and when academics lack the means to fully inspect or reproduce them — that claim becomes harder to sustain. The risk is not that universities cease producing research, but that they shift from being originators of foundational knowledge to mere consumers of systems designed elsewhere.

This shift has internal consequences as well. The spectre of deskilling, described earlier, becomes more acute if early-career researchers are trained primarily in prompting black-box systems rather than in mastering methods or data. Senior academics, meanwhile, may become dependent on corporate infrastructure that they cannot shape. Over time, this can diminish the craft and the pleasure of research, turning scholars from principal investigators into sophisticated users of proprietary tools. Again, the threat is less redundancy than irrelevance: a narrowing of the intellectual space in which academic judgment matters.

These challenges are not insurmountable. National and regional initiatives such as the US National Artificial Intelligence Research Resource (NAIRR) pilot and the Genesis Mission launched in late 2025, the United Kingdom's Isambard-AI facility, and European programmes built around EuroHPC and LUMI signal a recognition that shared public compute is essential to preserving an open research ecosystem (Jones, 2023; Taborsky, et al., 2025). Open-source consortia and public–private partnerships also demonstrate that meaningful collaboration is possible where governance safeguards academic independence.

Even so, the gravitational pull of the private sector remains strong. Unless universities articulate — and inhabit — roles that complement rather than mimic corporate laboratories, they risk being elbowed to the margins of the research landscape. In addition to reducing their legitimacy, this would almost certainly reduce their value in terms of the economic calculation of public funders. Recent experience in the United States has demonstrated that research funding can be a weapon to ensure ideological compliance of tertiary institutions; it is possible that other governments will use the distribution of research funding as a lever to discipline, reshape, or sideline institutions they regard as insufficiently aligned with national priorities. A university that cannot demonstrate a distinctive public value — in an area not easily provided by industry — will find it harder to resist such pressures.

The solution is unlikely to lie in competing on speed or scale. Rather, it may lie in reaffirming a different mission: pursuing inconvenient questions, sustaining epistemic diversity, and serving as custodians of trustworthy knowledge. The next section considers what such roles might entail.

# 4   New Roles for Universities?

If universities can no longer assume primacy in the production of knowledge across a wide range of domains — whether because industry now leads the research frontier or because AI automates tasks that once defined scholarship — the question becomes whether they have a role beyond nostalgic attachment to a fading era. The answer may lie not in competing with industry, but in reasserting forms of judgment, curation, and public service that private actors are poorly placed to provide. It turns out that a technology capable of producing information at unprecedented scale also enhances the value of institutions that can distinguish the signal from the noise.

## 4.1   Judgment

That role becomes especially important in the evaluation of research. For years, universities have drifted towards mechanized assessments of research quality: counting papers, tallying citations, ranking journals (Sugimoto and LarivìEre, 2018: pp. 127–33). Such proxies were always crude; in a world where generative systems can inflate publication volume at negligible cost, they become meaningless. Early indicators suggest that AI-assisted writing is already saturating preprint servers and peer-review workflows. Faced with this abundance, universities may rediscover what always distinguished them: not the ability to produce information, but the capacity to judge its worth.

International initiatives such as the Declaration on Research Assessment and the Leiden Manifesto offer pathways back to more qualitative, context-sensitive evaluation (Cagan, 2013; Hicks, et al., 2015; Hurrell, 2023). Their calls to decouple hiring and promotion from journal prestige were once seen as idealistic; they now appear prescient. Institutions adopting narrative CVs, for example, ask researchers to explain why a contribution matters to them personally and professionally — questions AI may not be able to answer (easily) on their behalf (Bordignon, et al., 2023). Such reforms nudge universities away from the metrics that machines can game and towards the human work of discerning quality, originality, and integrity. In practice, this implies hiring and promotion criteria that reward robustness, replication, and substantive contribution, rather than the sheer volume of AI-assisted outputs.

## 4.2   Curation

A second opportunity lies in the curation of trustworthy knowledge. Generative AI has made it trivial to produce polished but unreliable content: fluent text with no underlying argument, synthetic data falsely presented as empirical, or summaries of sources that no one has actually read (if they exist at all). Provenance becomes increasingly important when outputs can be generated by models trained on mixtures of copyrighted material, synthetic artefacts, and web-scraped text of uncertain quality. In this environment, universities can strengthen their role as custodians of the epistemic infrastructure: maintaining repositories that document where data came from, how models were trained, and what role AI played in analysis or writing.

Tools to support such work already exist. Model cards, dataset documentation standards, transparent reporting guidelines, and preregistration protocols can help ensure that AI-mediated research remains accountable (Gebru, et al., 2020; Mitchell, et al., 2019). But these tools require institutional commitment. Earlier eras of scholarship created libraries, journals, and citation practices to stabilize the record of knowledge; universities today may need to build the provenance infrastructure that anchors AI-era research in verifiable human practice. This extends outward as well. With public trust in expertise eroding, universities can play a civic role by modelling transparency, offering guidance on evaluating AI-generated information, and partnering with civic institutions to support epistemic literacy. They cannot police the entire information ecosystem, but they can illuminate parts of it.

A related role concerns the infrastructure of research itself. National efforts such as NAIRR represent early attempts to provide shared computation for public-interest research. Universities are well placed not only to use such facilities, but to help shape their governance. Academic norms emphasizing openness, replicability, and scrutiny can influence how shared compute is allocated, documented, and audited. To complement this, universities can strengthen their own commitments to open science: mandating persistent identifiers for datasets, enforcing FAIR data principles (Boeckhout, et al., 2018), requiring code and data release except where prohibited, and ensuring AI-use statements accompany research outputs. These measures do not guarantee perfect transparency, but they help maintain a research environment in which claims can be traced, tested, and reproduced.

## 4.3   Counterweight

Universities also have a distinctive role as ethical counterweights. Corporate laboratories excel at engineering and deployment, but their incentives do not always align with the broader values of science. Concerns about bias, fairness, safety, and long-term societal impact may be acknowledged, but are rarely prioritized. Universities, by contrast, are home to interdisciplinary expertise capable of examining how AI reshapes research norms — not only in terms of accuracy, but in relation to power, equity, and accountability. Their remit can be expanded to include the integrity of scientific workflows: independent audits of models used in research, field-specific guidance on appropriate use of generative systems, and scholarly scrutiny of claims made by industry. If there is a need for publicly accountable institutions to evaluate foundation models used in science, universities would be the obvious hosts for such work.

Perhaps the most important role, however, concerns training. If AI systems handle much of the routine labour of literature review, coding, or experimental design, universities must ensure that early-career researchers still acquire the conceptual and interpretive skills that underpin independent scholarship. Prohibiting AI use is unrealistic; the challenge is to structure training so that AI augments rather than replaces human thought. Some doctoral programmes are beginning to require researchers to document their use of generative tools, encouraging reflective practice rather than unthinking delegation (Boyd and Harding, 2025). Others are redesigning curricula to emphasize study design, uncertainty quantification, error analysis, and methodological reasoning — tasks unlikely to be fully automated, and central to the craft of research.

These skills also reflect what employers increasingly seek in a labour market transformed by AI. Surveys routinely highlight critical thinking, ethical judgment, communication, and problem framing as essential capabilities. Universities can therefore present themselves not simply as providers of technical training — which others may do more cheaply — but as institutions that cultivate the capacities necessary to navigate complex, ambiguous, and contested information environments.

None of this can be achieved by individual institutions alone. No university, however well-funded, can match the computational capacity of the largest firms. But consortia might. Regional alliances, disciplinary networks, and thematic collaborations can pool compute, data, and expertise to pursue research questions that industry is unlikely to tackle. The Masakhane initiative in African NLP offers a compelling model: a distributed network of scholars addressing neglected problems through shared expertise rather than capital. Universities in the Global South, especially, may find such collaborations essential to asserting agency in a research ecosystem otherwise shaped elsewhere.

# 5 Conclusion

The disruption of research by AI is more than a technical challenge. Beyond questions of productivity and discovery lies a deeper concern: the meaning (and pleasure) of academic life itself. For many scholars, writing and experimentation are not simply the elaboration of ideas already formed, but ways of thinking them into existence. If those processes are displaced by machines, there is a risk that academics cede epistemic agency, becoming overseers of outputs they did not meaningfully shape. That would not merely alter research methods; it would diminish the vocation that attracted professors to scholarship in the first place.

AI has accelerated discovery while shifting its locus towards private laboratories and saturating the scholarly record with synthetic outputs. Universities may soon be unable to claim primacy in discovery, but they can still sustain the conditions under which discovery is trustworthy and publicly valuable. As argued at the outset, academic authority has never rested solely on the production of new information, but on the capacity of universities to sustain epistemic practices — judgement, transparency, reproducibility, and independence from concentrated power — through which knowledge claims acquire credibility beyond the individuals who advance them. The roles available to universities — as evaluators, curators, and public counterweights to private interests — may be less prominent than in the glory days of academic primacy. They may also be more important. In a world awash with information but starved of discernment, universities may find renewed purpose not by winning the race for scale, but by helping society decide what is worth knowing.

For the larger risk is that expertise itself is devalued. If universities fail to sustain academic authority — understood not as prestige or output, but as the institutional capacity to render knowledge credible and contestable — societies may increasingly turn to AI systems as substitutes for human judgement rather than as tools to augment it. Knowledge would still be produced, but without shared standards for validation, provenance, or accountability, leaving claims to be assessed by fluency, scale, or commercial utility rather than by epistemic warrant or public value. In such a world, universities might persist as organizations, but no longer as epistemic institutions: their role reduced to training users of proprietary systems and legitimizing outputs they can neither fully interrogate nor reproduce. The ultimate question, then, is not whether universities can adapt to artificial intelligence, but whether knowledge can retain authority if the institutions that historically validated it do not.